# Methodical engineering of defects in $Mn_xZn_{1-x}O$ (x = 0.03, and 0.05) nanostructures by electron beam for nonlinear optical applications: A new insight

Albin Antony[a], P. Poornesh[a,*], I.V. Kityk[b], K. Ozga[b], J. Jedryka[b], Reji Philip[c], Ganesh Sanjeev[d], Vikash Chandra Petwal[e], Vijay Pal Verma[e], Jishnu Dwivedi[e] a *Department of Physics, Manipal Institute of Technology, Manipal Academy of Higher Education, Manipal, Karnataka, 576104, India*
b *Institute of Optoelectronics and Measuring Systems, Faculty of Electrical Engineering, Czestochowa University of Technology, Armii Krajowej 17, PL-42- 201, Czestochowa, Poland*
c *Light and Matter Physics, Raman Research Institute, C. V. Raman Avenue, Sadashivanagar, Bengaluru, 560 080, India*
d *Department of Physics, Mangalore University, Mangalore, Karnataka, 574199, India*
e *Industrial Accelerator Section, PSIAD, Raja Ramanna Centre for Advanced Technology, Indore, 452012, M.P, India*

ABSTRACT

A series of $Mn_xZn_{1-x}O$ (x = 0.03, 0.05) nanostructures have been grown via the solution based chemical spray pyrolysis technique. Electron beam induced modifications on structural, linear and nonlinear optical and surface morphological properties have been studied and elaborated. GXRD (glancing angle X-ray diffraction) patterns show sharp diffraction peaks matching with the hexagonal wurtzite structure of ZnO thin films. The upsurge in e-beam dosage resulted in the shifting of XRD peaks (101) and (002) towards lower angle side, and increase in FWHM value. Gaussian deconvolution on PL spectra reveals the quenching of defect centers, implying the role of electron beam irradiation regulating luminescence and defect centers in the nanostructures. Irradiation induced spatial confinement and phonon localization effects have been observed in the films via micro Raman studies. The later are evident from spectral peak shifts and broadening. Detailed investigations on the effect of electron beam irradiation on third order nonlinear optical properties under continuous and pulsed mode of laser op-eration regimes are deliberated. Third order absorptive nonlinearity of the nanostructures evaluated using the open aperture Z-scan technique in both continuous and pulsed laser regimes shows strong nonlinear absorption coefficient $\beta_{eff}$ of the order $10^{-4}$ cm/W confirming their suitability for passive optical limiting applications under intense radiation environments. Laser induced third harmonic generation (LITHG) experiment results supports the significant variation in nonlinearities upon electron beam irradiation, and the effect can be utilized for frequency conversion mechanisms in high power laser sources and UV light emitters.

## 1. Introduction

In the past few decades wide band gap semiconductors (WBG) materials have attracted much interest due to its applications in optics and solid state electronics [1–3]. Among the widely investigated WBG semiconductor materials ZnO belonging to II-VI group with energy band gap of 3.36eV has gained a lot of studies due to possible applicability in various fields especially as a perspective material for ultrafast UV light modulators and near-UV emitters [1–6]. The devices based on nonlinear optical effects are at pressing need. The available conventional nonlinear optical materials have reached its technical limit. Furthermore there is a demand for nanostructured nonlinear optical materials which can be easily integrated into nonlinear optical devices [6–11]. It's important for future optoelectronics and photonics to search and design of new nonlinear optical materials with large nonlinear optical responses [6–10]. These new materials seems to be promising and enable novel devices with high performance and new functionalities like reduced cost of production, easily downscaling to industry standards and enhanced performance etc... Studies related to the nanostructured ZnO thin films [12,13] pointed out that this material exhibits a strong and efficient second and third harmonic generations. But in order to meet the requirements of cutting-edge nonlinear optical devices with reduced cost and better-quality performance, the traditional physical and chemical properties and method of synthesis of ZnO nanostructures should be modified.

In the present study host ZnO lattice was doped by Mn with different concentration. Mn has the advantage because of its half-filled 3d orbitals which aids its incorporation into the ZnO lattice and relatively comparable ionic radii between $Zn^{2+}$ and $Mn^{2+}$ [14]. Another point to be highlighted is the method of thin film deposition and choice of substrate. Different method of thin film synthesis led to different growth behavior, hence different physical properties and thus nonlinear behavior of the samples can also be varied. In the present study we employed solution based air assisted spray pyrolysis technique [15] for the growth of $Mn_xZn_{1-x}O$ nanofilms. It is a cost-effective, scalable and well explored method to grow $Mn_xZn_{1-x}O$ nanostructures from aqu-eous salt solutions of Mn and Zn. It allows a large range of diverse geometries by wavering the films growth conditions [16]. Furthermore the chemical and physical properties of the deposited films can be tailored by imposing various external conditions such as substrates, annealing temperature, strain, electric and magnetic field, electron and ion beam irradiations [17,18] etc. Up today there are no reports con-cerning the influence of energetic electron beam irradiation on tailoring the nonlinear optical properties of $Mn_xZn_{1-x}O$ nanostructures. There-fore, investigating irradiation induced effects and subsequently en-hancing or modulating the physical and chemical properties of $Mn_xZn_{1-x}O$ thin film is significant to understand and further modification of the materials for nonlinear optical device applications such as frequency converters, optical limiters, modulatos, triggers etc...Herein the effect of 8 MeV electron beam irradiation on third order optical suscep-tibilities for $Mn_xZn_{1-x}O$ nanostructures were studied and elaborated for the first time. The electron beam induced modifications in linear op-tical, morphological, and structural properties of the films were studied via Glancing angle X-ray diffraction(GXRD), Atomic force microscopy (AFM), Photoluminescence and Raman spectroscopy.

## 2. Experimental details

### 2.1. Thin film preparation

To grow $Mn_x Zn_{1-x}O$ thin nanocrystalline films with Mn con-centration(X) of 3% and 5% we implemented a cost effective spray pyrolysis film growth technique [15]. Appropriate proportions of powders of Zinc chloride and Manganese chloride tetra hydrate were dissolved in a 1:3 solution of ethanol and double distilled water. The molar content of the solutions were fixed to be 0.05M. The prepared Mn dopant solution were added to host solution by prefixing the dopant concentration as 3% and 5%. The deposition parameters followed for the growth of $Mn_X Zn_{1-X} O$ thin films were reported elsewhere [15].

### 2.2. Characterization

The physical properties such as structural, linear optical and mor-phological features for deposited films were studied using various char-acterization tools. The morphological changes in the $Mn_xZn_{1-x}O$ (x = 0.03, 0.05) thin nanofilms were studied using Bruker Icon Atomic force microscope (AFM). The variations of the topological parameters were done using Bruker Nanoscope analysis software. Electron beam simulated with respect to crystallinity, orientation and other structural parameters were analyzed by X-ray diffractometer (GXRD) within a 2Θ range of $20^0$-$80^0$. Shimadzu 1800 UV-Visible spectrophotometer were applied for linear absorption spectra measurements. The non-radiative and radiative recombination in the grown films were investigated by Raman spectroscopy (Horiba JOBINYVON LabRAM HR) and Photoluminescence spectroscopy (Horibafluromax-4 Spectrofluorometer).

### 2.3. Open aperture Z-scan technique

Open aperture Z-scan measurement were applied to study the third order nonlinearity of the deposited films [19,20]. The measurements have been executed both in continuous as well as pulsed laser regimes. The excitation source used for the measurement in continuous wave regime was He-Ne laser with a wavelength of 632.8 nm and input power of 20 mW. The pulsed regime measurements were carried out using Continuum Minilite IQ-switched Nd: YAG laser with nanosecond pulses (532 nm, pulse energy 100 µJ, pulse duration 5ns, pulse fre-quency repletion 10Hz).

### 2.4. Third harmonic generation

An Nd-YAG fundamental laser with 8 ns pulse duration, 1064 nm wavelength and 10 Hz pulse frequency repletion rate were employed as the fundamental beam and the photo induction to the samples were done by modulated cw 532 nm CW laser [21]. The studied samples were put on a rotating table. Levels of obtained fundamental and third harmonic signals were measured using a Tektronix MSO 3054 oscillo-scope with sampling of 2.5 GS. The obtained data points were averaged around 400 points throughout the film surfaces.

### 2.5. Electron beam irradiation

Electron beam line from linear accelerator (LINAC) at RRCAT Indore, India were employed for the electron beam irradiation experi-ments. The $Mn_xZn_{1-x}O$ (X = 0.03, 0.05) thin films were irradiated at predetermined dosages starting from 5 kGy to 20 kGy in steps of 5. The energy of the electron beam line was fixed to 8 MeV. The time of beam exposure to samples were designed based on the dosage to be delivered.

## 3. Results and discussion

### 3.1. Influence of electron beam treatment on linear optical properties

Linear absorption spectra of $Mn_xZn_{1-x}O$ (X = 0.03, 0.05) thin films were noted within the wavelength range varying within 350 nm-1000 nm and are depicted in Fig. 1 for different irradiation dosage. The increase in the optical scattering loss of light from the grain boundaries favours an enhancement of the absorption as evident from figure. The aggregation of the grains in the irradiated films will effect in the for-mation of large grain boundaries enhancing the optical scattering ef-fects [22]. The red spectral shift observed in the absorption edge of the films results in the shrinkage of band gap. The band gap of the irra-diated films were found out by implementation of Tauc plot method [23] and is shown in Fig. 2. The obtained band gap energy values are depicted in Table 1. The formation of localized defect states upon electron beam irradiation results in the decrease of the band gap energy values. The more discussion on the formation of localized defect states were done in Sec 3.3.

### 3.2. Electron beam irradiation effects on structural and surface morphological properties

GXRD pattern are obtained using Cu Kα radiation source with λ = 1.54$A^0$ and shown in Fig. 3. The results are used to study the effect of irradiation on crystallographic structure, nanocrystalline size, lattice strain and bond length. GXRD pattern showing sharp diffraction peaks and they are matched with hexagonal wurtzite structure of ZnO thin films (JCPDS-036-1451). The observed XRD diffraction peaks in the films are indexed to (002), (101), (102), (110), (103), (112), and (201) with preferred growth orientation along (002) C-axis direction. The angle shift of the diffraction peak position observed with irradiation dosages are attributed to elongation of lattice parameters which results in compressive strain in the nanostructures [17,18]. The absence of Mn related peaks in the GXRD pattern may indicate incorporation of Mn into ZnO lattice due to substitution of Zn atoms position or in other way the impurity content of Mn in the films are lower than the detection limit of GXRD used. The changes noticed in the lattice parameters are also an indication of expansion of crystal lattice. Enhancement in the intensity of diffraction peak corresponds to the influence of irradiated electron beam on the scattering factors of $Mn_xZn_{1-x}O$ unit cell [17]. The discrepancies observed in the intensity of diffraction peaks fall out the variations in the crystalline origin of the films on irradiation.

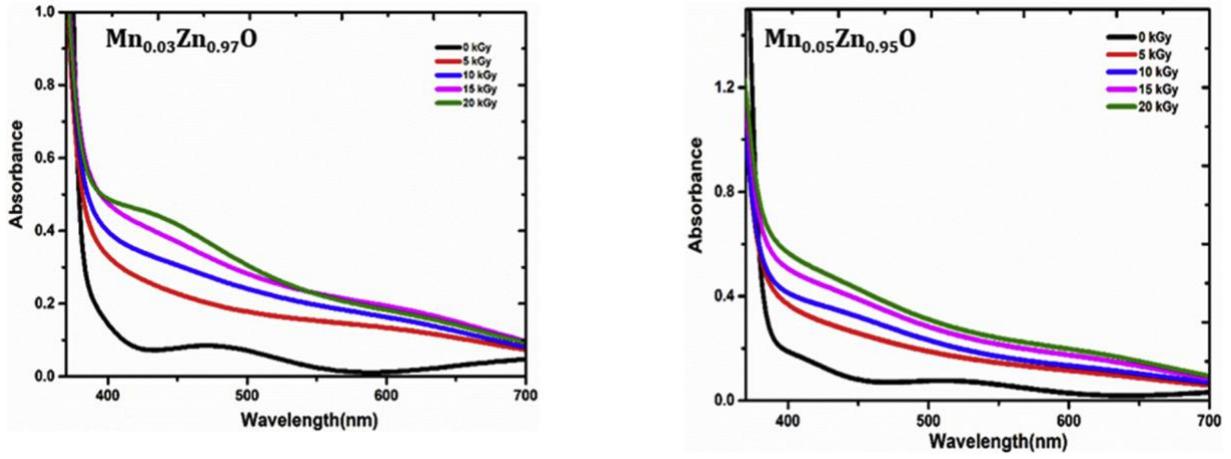

**Fig. 1.** Absorption spectra of $Mn_XZn_{1-X}O$ (x = 0.03, 0.05).

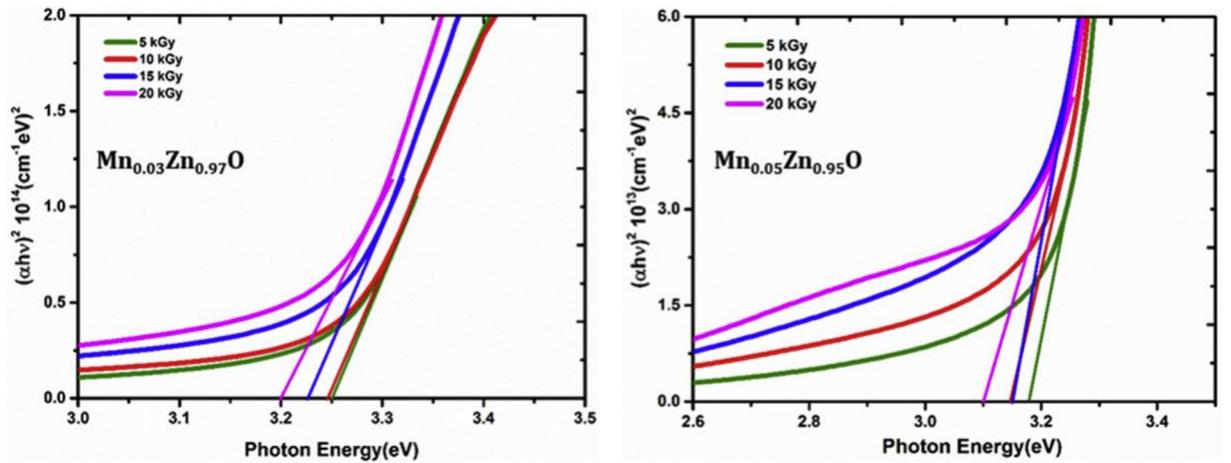

**Fig. 2.** Tauc plot of $Mn_XZn_{1-X}O$ (x = 0.03, 0.05).

The electron beam induced variations on structural parameters were found out and depicted in Table 2. The peak at $2\Theta \sim 34.6^0$ is observed to be strongest for all films except for pristine $Mn_{0.03}Zn_{0.97}O$ which shows at $2\Theta \sim 36.6_0$. Crystallite size, D of the films were evaluated using Debye Scherer formula reported elsewhere [22].

The irradiation effects results in the decrease in the crystallite size as observed from Table 2. The decrement of crystallite size owes to the decrease in the grain size or deterioration of crystallinity. The decre-ment in the crystallinity or increase in the FWHM can be attributed to the incorporation of lattice defects in the nanostructure by e-beam ir-radiation [17,24]. The broadening of the x-ray diffraction peak which results in the decrease in the crystallite size further associated with the formation of localized strain in the lattice. These arised strains in the films are due to the creation of extended defects or point defects which in turn results in the systematic modifications of atoms positions from their original position [24]. Close inspection of Fig. 3 indicates a shift in the diffraction angle upon irradiation. The prominent diffraction along (101) and (002) plane exhibits a lower angle shift on higher dosages as evident from Fig. 4. Irradiation induced elongation in the lattice parameters which results in the elongation of bond length can be attributed to the observed shift in the diffraction peak towards lower $2\Theta$. Electron beam induced irregularity in the crystal structure or the amount of defects induced in the lattice can be explained by finding out the dislocation density of the nanostructure. Dislocation density represents the magnitude of defects in the samples which can be find out using equation (1) and depicted in Table 2

$$\delta = \frac{1}{D^2} \tag{1}$$

Where D is nanocrystallite size. The exposure to energetic e-beam resulted in enhancement of dislocation density underlying the enhancement in lattice defects. It is noteworthy that with the upsurge in e-beam dosage, the XRD peaks (101) and (002) are shifted toward lower angle side, and its FWHM is increased. But the basic crystal structure of $Mn_xZn_{1-x}O$ nanostructure against electron beam irradiation remains same indicates the good structural stability.

In order to determine the changes in surface morphology under

**Table 1**
Variation in energy band gap at different irradiation dosage.

| Dosage (kGy) | $Mn_{0.03} Zn_{0.97} O$ (eV) | $Mn_{0.05} Zn_{0.95} O$ (eV) |
|---|---|---|
| 0 | 3.26 | 3.25 |
| 5 | 3.17 | 3.25 |
| 10 | 3.14 | 3.24 |
| 15 | 3.11 | 3.22 |
| 20 | 3.15 | 3.19 |

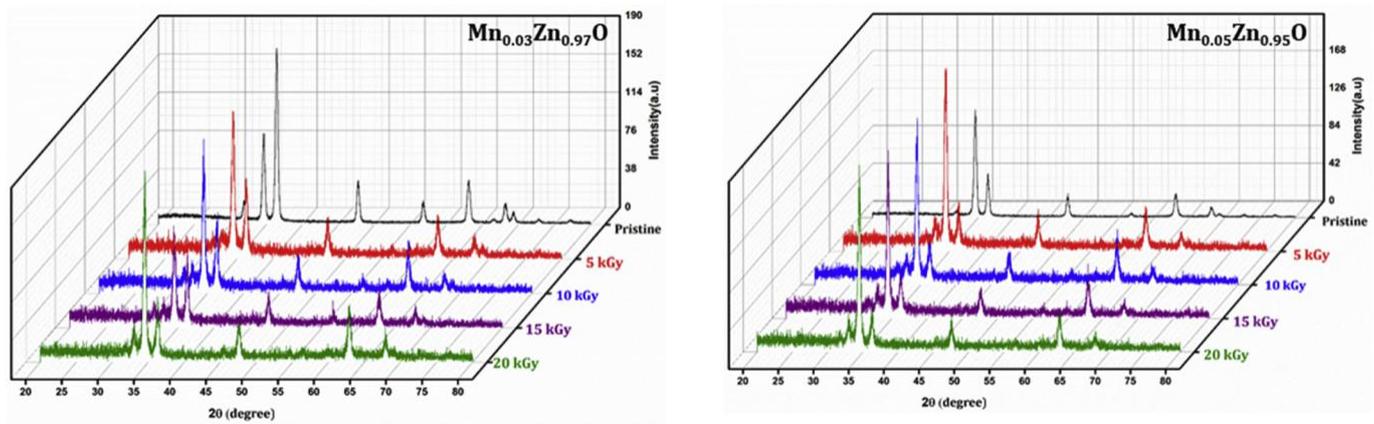

**Fig. 3.** 3D GXRD diffractogram of $Mn_xZn_{1-x}O$ (x = 0.03, 0.05).

**Table 2**
Variation in structural parameters upon electron beam irradiation.

$Mn_{0.03}Zn_{0.97}O$

| Dosages(kGy) | 'c'(002) Å | 'a'(101) Å | Bond Length(L) Å | Crystalline size(D) nm | Strain ($10^{-3}$) | Dislocation density($10^{15}$) |
|---|---|---|---|---|---|---|
| 0 | 5.169 | 3.232 | 1.966 | 18.83 | 1.83 | 2.81 |
| 5 | 5.197 | 3.250 | 1.976 | 16.02 | 2.16 | 3.90 |
| 10 | 5.199 | 3.227 | 1.967 | 16.62 | 2.08 | 3.61 |
| 15 | 5.206 | 3.240 | 1.973 | 16.24 | 2.13 | 3.79 |
| 20 | 5.190 | 3.231 | 1.968 | 16.60 | 2.09 | 3.60 |

$Mn_{0.05}Zn_{0.95}O$

| Dosages(kGy) | 'c'(002) Å | 'a'(101) Å | Bond Length(L) Å | Crystalline size(D) nm | Strain ($10^{-3}$) | Dislocation density($10^{15}$) |
|---|---|---|---|---|---|---|
| 0 | 5.188 | 3.238 | 1.970 | 18.00 | 1.92 | 3.07 |
| 5 | 5.195 | 3.230 | 1.968 | 17.00 | 2.03 | 3.45 |
| 10 | 5.188 | 3.238 | 1.970 | 17.10 | 2.02 | 3.41 |
| 15 | 5.199 | 3.247 | 1.975 | 16.74 | 2.07 | 3.60 |
| 20 | 5.197 | 3.239 | 1.972 | 17.42 | 1.99 | 3.30 |

influence of e-beam irradiation atomic force microscopy (AFM) technique was employed. The films were scanned in tapping mode configuration at a scan area of 5μm×5 μm. The 3D AFM images were recorded and analyzed by the aid of Nanoscope analysis software. The changes observed in the average surface roughness of the films were obtained and tabulated in Table 3. In comparison to non-irradiated $Mn_xZn_{1-x}O$ the e-beam irradiated films demonstrate a variation in surface morphology clearly observed in Fig. 5(a–c). The Root mean square roughness (RMS) value shows an enhancement upon irradiation due to the formation of defects in the sample. It is worthy to be emphasized that e-beam irradiation stimulated large reduction of agglomeration sites which further leads to fragmentation of larger grains into smaller ones with more ordered and shaped grain structures [25]. The changes in shape and size of the grains at higher electron beam dosages are due to the high density of radiative and non-radiative defects induced by the irradiation. There is no any substantial changes observed in the topo-graphical features especially in the average surface roughness and apart from that higher damage threshold shown by the films leads to the credibility of investigated material in nonlinear device application.

### 3.3. Electron beam induced effects investigated at ambient temperature. Photoluminescence (PL) spectroscopy

Apart from the optical properties PL studies will give an information about structural defects such as Zinc and oxygen interstitials, oxygen vacancy, Zinc vacancy and other surface properties The PL spectra of non-radiated and e-beam irradiated $Mn_xZn_{1-x}O$ (x = 0.01,0.03,0.05) nano-structures were recorded at an excitation wavelength of 325 nm and are shown in Fig. 6(a–c). Gaussian deconvolution fitting is performed for obtained PL spectra of the nanostructures in order to identify the radiative transitions in the films reflecting different native defect centers. The fitting parameters used to perform the deconvolution was kept uniform for all the specimens. Fig. 6 a shows the PL spectra of pristine $Mn_xZn_{1-x}O$ (0.03, 0.05) films. The Gaussian fitting shows existence of luminescent centers which can be attributed to various defect centers presented in the films. A strong and narrow peaks observed at 3.11eV, 3.15 eV, 3.12 eV in the UV region can be endorsed to near band edge emission (NBE) possibly raised as a result of exciton recombination effects [26,27]. It is observed from the PL spectra that all the three investigated films exhibit strong visible light emission with different centers which can be recognized to manifestation of various structural defects. Violet emission observed around 2.96 eV, 2.85 eV, 2.94 eV were due to Zinc interstitial ($Zn_i$) defect center which traps the electrons while the de-exciting from conduction band to valence band [26–30]. A prominent and strong blue I and blue II emissions demonstrated by all the films is due to the presence of oxygen anti-site defect site ($O_{zn}$) in the films [27]. The peaks observed at 2.32 eV, 2.38eV, 2.37 eV were attributed to green luminescence emission from the nanostructures. Green emission is assigned to oxygen vacancy defects of the titled the films which give rise to the recombination of photo generated holes and electrons. The same observation has been reported in different references [26–30]. The yellow emission peaks observed at 2.19 eV and 2.18 eV in the investigated films are assigned to the presence of absorbed hydroxyl group (OH) and interstitial oxygen ($O_i$) [27].

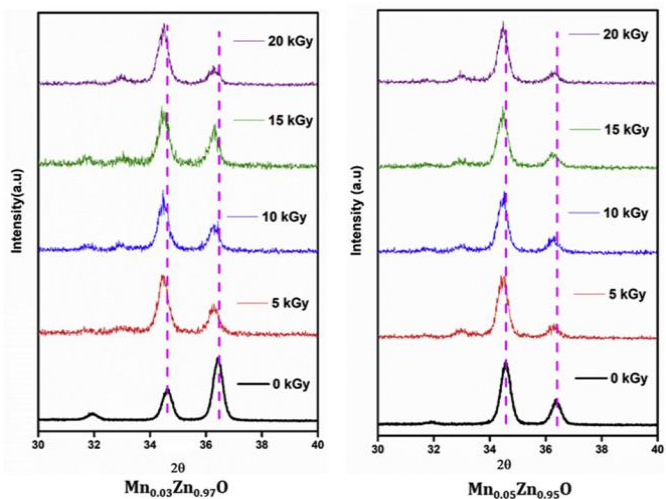

**Fig. 4.** XRD peak shift upon e-beam irradiation.

**Table 3**
Variation in average surface roughness upon electron beam irradiation.

| Dosage (kGy) | Surface Roughness (nm) | |
| --- | --- | --- |
| | $Mn_{0.03}Zn_{0.97}O$ | $Mn_{0.05}Zn_{0.95}O$ |
| 0 | 13.2 | 19.5 |
| 5 | 14.6 | 28.6 |
| 10 | 11.8 | 20 |
| 15 | 10.4 | 19.4 |
| 20 | 16.9 | 22 |

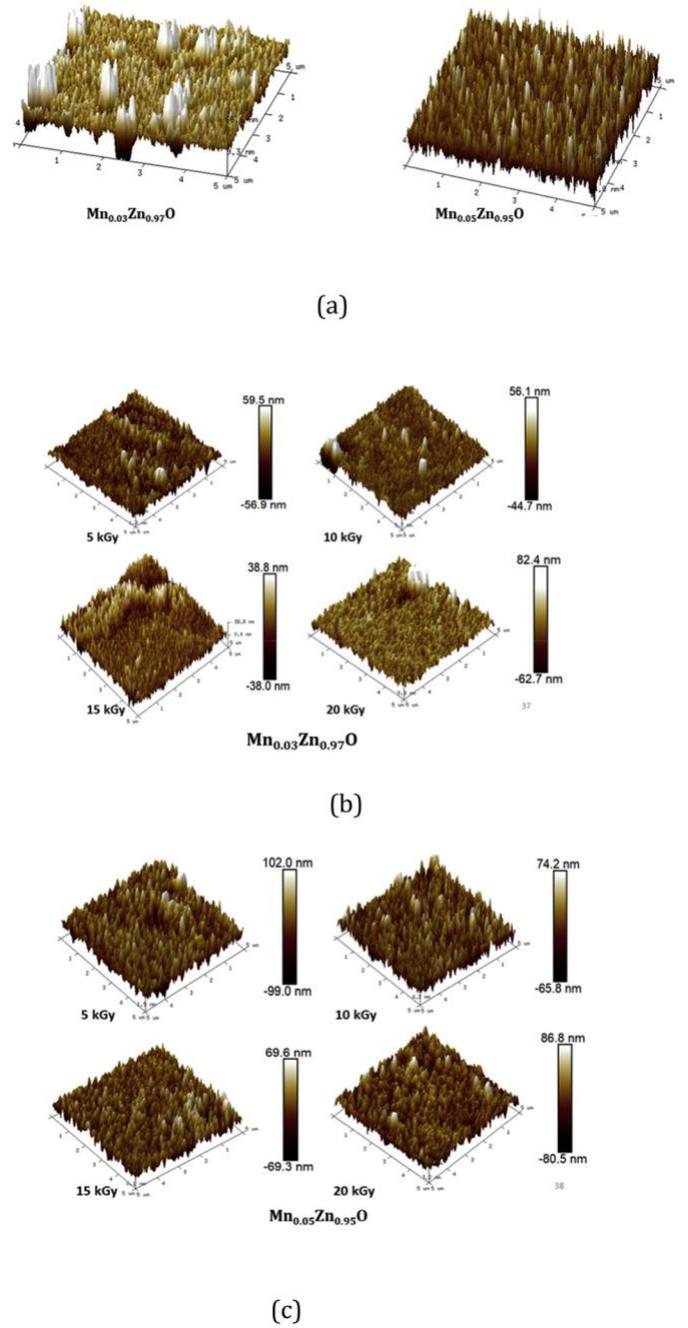

The PL spectra of electron beam irradiated $Mn_xZn_{1-x}O$ (x = 0.03, 0.05) films have been mainly dominated by four luminescent spectral peaks situated around~ 2.49 eV, 2.63 eV, 2.77 eV and 2.91 eV which can be formed by four color centers such as blue-green, blue, violet-II and violet, respectively. Fig. 6(b) and c shows the deconvuluted PL spectra of irradiated nanostructures. The four visible emission centers corresponds to defect center transition and inter-band carrier transitions. The electron transition probabilities between $Zn_i$ site and valence band are enhanced due to the observed violet emission around 2.91 eV. A prominent Violet II peak emission observed at 2.77eV and 2.78eV were consigned to the transfer of electron between the $Zn_i +$ center and valence band [27,29]. Blue emission peak observed in the films were attributed to presence of oxygen anti-site defect centers for the nanos-tructures [27]. The non-equilibrium condition like high energy irra-diation with ions or electron is the required environment for the $O_{Zn}$ defect sites to induce in the materials which are likely present in the current investigation. Blue-green emission peak observed in the nanostructures can be attributed to the presence of intrinsic oxygen vacancy defect sites. Moreover, it is observed a substantial decrease in the PL emission intensity upon e-beam incorporation. This behavior can be due to the formation of localized surface states below the conduction band. As a result of these surface states, non-radiative transition of the carriers is enhanced occurring from the conduction band to the surface states below the conduction band. The saturation of PL emission centres observed in irradiated PL spectra can be probably due to the re-combination of $Zn_i$ and $V_{Zn}$ defect sites. As a consequence a spatial separation between these defects sites will reduce on irradiation which aids to recombination mechanism. This also indicates the saturation of defect concentration in the films upon electron beam irradiation [31].

*3.4. Electron beam induced effects investigated by Raman spectroscopy*

The non-radiative transitions and structural disorder in $Mn_xZn_{1-x}O$ (x = 0.03, 0.05) thin films were investigated by Raman spectroscopy in back scattering geometry. The excitation source used was 532 nm continuous wave laser in $180^0$ orientation. The assignment of phonon modes were done by group theory analysis predicted based on wurtzite ZnO nanostructures [32]. Fig. 7 (a) and 7(b) shows the Raman spectra of $Mn_xZn_{1-x}O$ (x = 0.03, 0.05) thin nano-films. It is observed that spectra was dominated by two spectral peaks at 330 and 437 $cm^{-1}$. The mode at 437 $cm^{-1}$ were assigned to E2 (higher energy) mode, and the modes at 330 $cm^{-1}$ assigned as E2 (higher)–E2 (lower energy) modes, respectively [32–34]. From Fig. 7 it is clear that the Raman peaks of $Mn_x Zn_{1-x}O$ (x = 0.03, 0.05) correspond to those of ZnO, except in few which is evident from the spectral images. It is observed from the spectra that the intensity of the dominant peak at 437 $cm^{-1}$ mode is changed due to the fluctuation in the crystal upon doping and the Raman band at 600–700 $cm^{-1}$ becomes stronger upon increasing Mn concentrations [35]. The Raman modes assigned to $Mn_x Zn_{1-x} O$ (x = 0.03, 0.05) were tabulated in Table 4. The presence of narrow peak at 439 cm$^{-1}$ with E2H characteristic phonon modes were domi-nated in all deposited films is a signature peak of wurtzite ZnO. The variations in the intensity and peak shifting of this modes corresponds to the disordering degree in the films. The line shape of E2H mode at 439 cm$^{-1}$ depends on composition of Mn in the material system. The presence of an intense broad band around 560 cm$^{-1}$ with A1 (LO) vi-bration modes were observed in all doped films. The incorporation of Mn has increased the intensity of A1 (LO) mode which attributes to the presence of Mn in the films. It's an indication that Mn substitutes Zn in the lattice and 560 cm$^{-1}$ modes related to the local vibrations involving the Mn atom. The presence of 333 cm$^{-1}$ peak in all corresponds to the quality of the deposited samples [32–35].

**Fig. 5.** 3D AFM image of (a) unirradiated $Mn_xZn_{1-x}O$ (x = 0.03, 0.05) (b) $Mn_{0.03}Zn_{0.97}O$ (c) $Mn_{0.05}Zn_{0.95}O$ at different irradiation dosage.

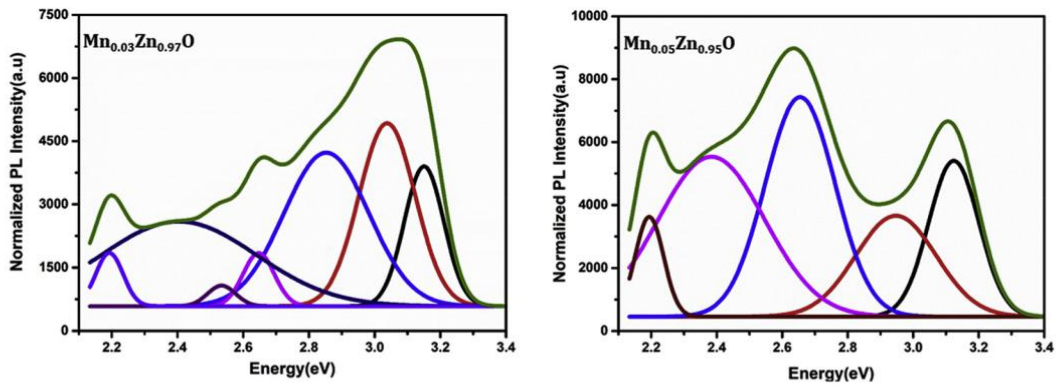

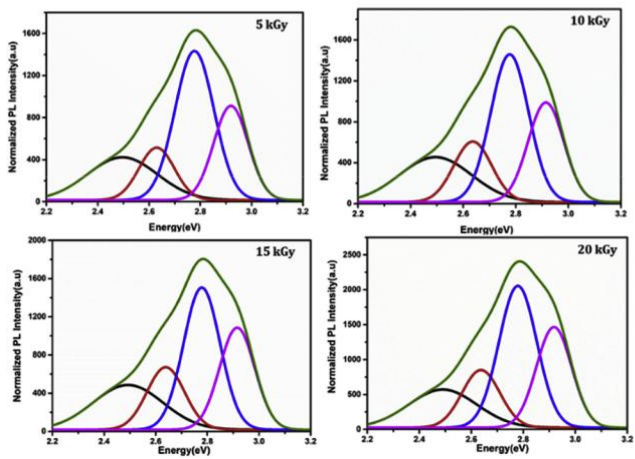

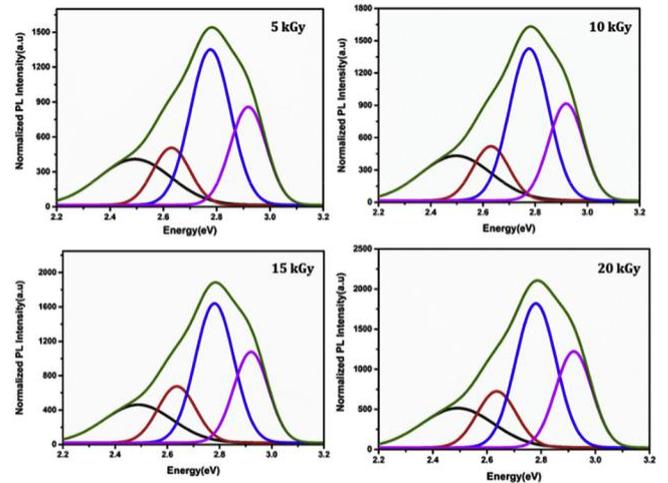

**Fig. 6.** Deconvoluted PL spectra of (a) pristine $Mn_xZn_{1-x}O$ (x = 0.03, 0.05) (b) $Mn_{0.03}Zn_{0.97}O$ (c) $Mn_{0.05}Zn_{0.95}O$ at different irradiation dosage.

Raman spectra was obtained for irradiated samples in order to study the formation of defect centers in the samples. The reports of various groups [36–39] predicts that there is a spectral shift in the Raman peaks and also distortions are observed when the samples are undergone ir-radiation treatment. For $Mn_x Zn_{1-x} O$ (x = 0.03, 0.05) films consider-able changes are observed upon irradiation. As evident from Fig. 7 the dominant peaks present in the films were assigned to E2H, E2H-E2L and A1 (LO) modes. It is noteworthy that upon e-beam exposure the ob-served Raman peaks underwent a shift in the spectral position with broadened peaks as compared pristine. Table 4 shows the peak posi-tions and corresponding phonon modes assigned. From Fig. 7(a) Raman spectra of $Mn_{0.03}Zn_{0.97}O$ thin films the E2H mode was observed at 439 cm$_{-1}$ for non-radiated samples and they were blue shifted and

A1(LO) mode at 557 cm$^{-1}$ were spectrally red shifted upon electron beam irradiation. The shifting of A1 (LO) phonons also corresponds to the slight variations in the preferred C axis growth orientation upon electron beam irradiation. A1 (LO) phonon can appear only when the c axis of wurtzite ZnO is parallel to the sample surface. The absence and reappearance of E2H-E2L mode at around 330 cm$^{-1}$ attributes to the changes in the samples crystallinity, crystalline size and strain occurred on electron beam treatment. The Raman spectra analysis of $Mn_{0.05} Zn_{0.95}O$ thin films also demonstrate identical behavior.

The asymmetric broadening and shifting of spectral peaks to lower and higher frequency sides observed in electron beam irradiated spectra can be likely due to three possible mechanisms arised in the films [36–41]. The presence of defect centers such as oxygen and Zinc va-cancy, zinc excess, anti-site oxygen and other surface impurities give rise to phonon localization effects. Furthermore spatial confinement effects induced in the samples can also be a possible explanation me-chanism. Electron beam induced heating in nanostructure and induc-tion of tensile strain in sample can also led to peak shifting process.

### 3.5. Electron beam irradiation effects on nonlinear optical properties

Electron beam induced modifications on third order nonlinear op-tical properties of $Mn_xZn_{1-x}O$ (x = 0.03, 0.05) nanostructures were studied using single beam Z-scan technique (both continuous and pulsed regime) and Third harmonic generation technique (THG).

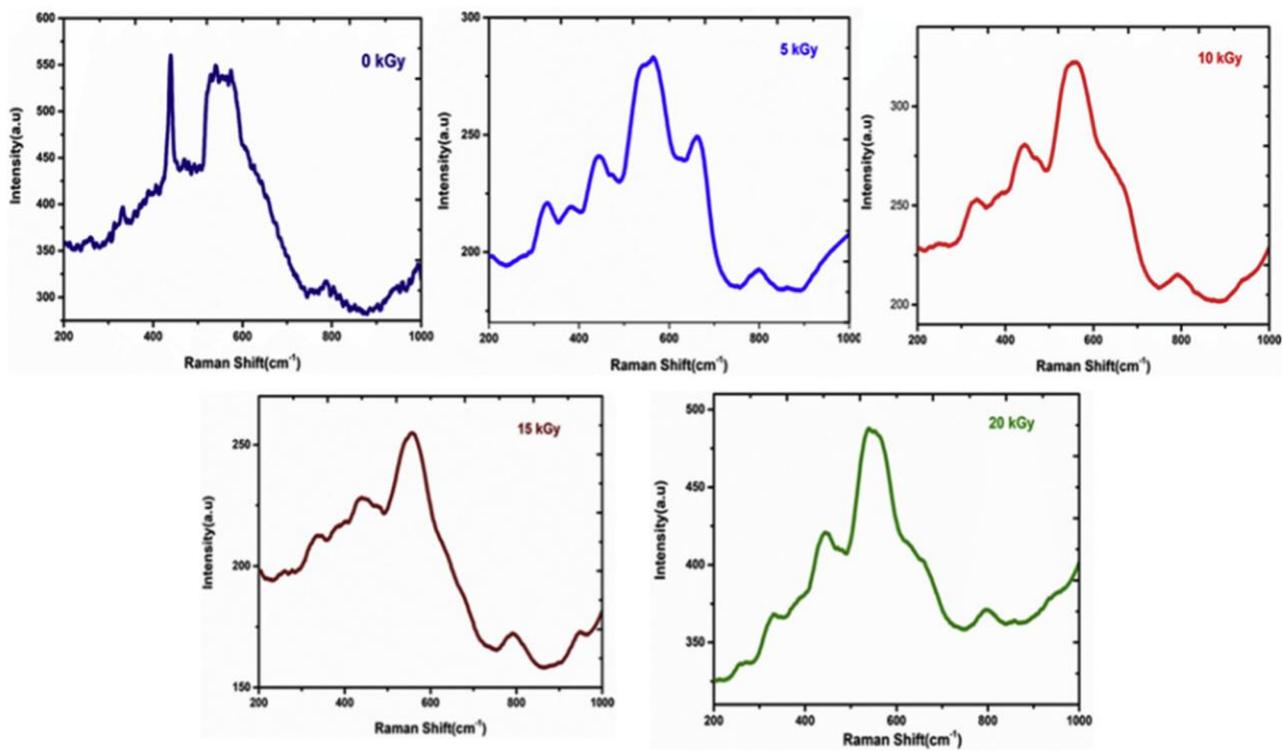

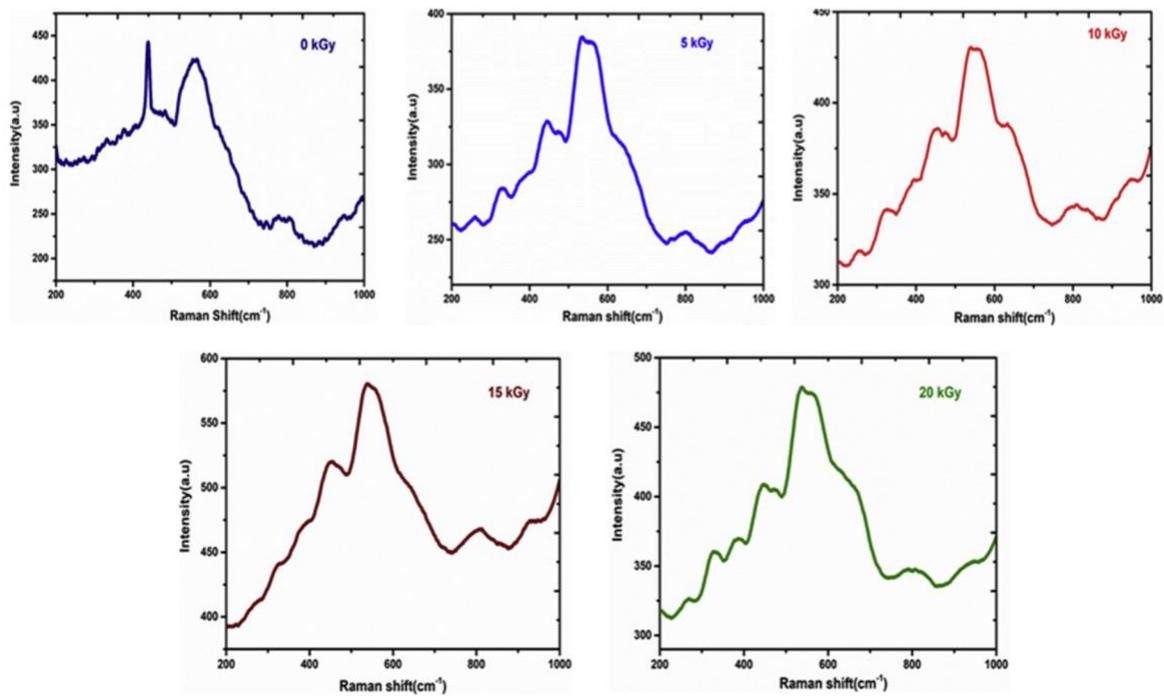

**Fig. 7.** Raman spectra of (a) $Mn_{0.03}Zn_{0.97}O$ (b) $Mn_{0.05}Zn_{0.95}O$ at different irradiation dosage.

**Table 4**

Spectral peak position and Phonon mode assignment on $Mn_xZn_{1-x}O$.

| Dopant Concentration | Phonon modes | Electron beam dosage (kGy) | | | | |
|---|---|---|---|---|---|---|
| | | 0 | 5 | 10 | 15 | 20 |
| $Mn_{0.03}Zn_{0.97}O$ | $E_{2High}$ mode(cm$^{-1}$) | 439 | 440 | 445 | 439 | 444 |
| | A1LO mode(cm$^{-1}$) | 557 | 554 | 551 | 560 | 539 |
| | E2High-E2Low mode(cm$^{-1}$) | - | - | 322 | 336 | 327 |
| | A1(TO)(cm$^{-1}$) | - | 379 | - | - | - |
| $Mn_{0.05}Zn_{0.95}O$ | $E_{2High}$ mode(cm$^{-1}$) | 439 | 441 | 444 | 448 | 441 |
| | A1LO mode(cm$^{-1}$) | 565 | 531 | 553 | 540 | 540 |
| | E2High-E2Low mode(cm$^{-1}$) | - | 330 | 322 | 322 | 330 |
| | A1(TO)(cm$^{-1}$) | - | - | 391 | 380 | 384 |

*3.5.1. Z-scan technique*

The Single beam open aperture Z-scan technique was employed to study the third order absorptive nonlinearity of the films [19,20]. The measurements have been done in both continuous and pulsed laser regimes. In both cases the nonlinearity arises essentially from the absorption of the excited states [21].

Fig. 8 (a) and 8(b) show open aperture Z-scan trace of $Mn_xZn_{1-x}O$ (x = 0.03, 0.05) nanostructures under cw laser excitation. The sig-nature trace obtained from the measurement can be attributed pre-vailingly to two different phenomena. The trace which unveils a valley at the focus corresponds to reverse saturable absorption mechanism (RSA) which recognize as positive absorption nonlinearity and the trace which unveils a peak at the focus attributes negative absorption non-linearity which arised due to saurable absorption (SA) mechanism. The obtained open aperture Z-scan trace shows that e-beam exposure results in a switching over of the nonlinearity from RSA to SA and vice versa. The presence of RSA in semiconductor nanostructures can be explained by different nonlinear processes such as two photon absorption (TPA), excited state absorption (ESA), induced scattering, free carrier absorp-tion (FCA)etc. or a blend of any of these processes [42,43]. In view of the energetics required for TPA process to occur, the energy of the excitation source is 1.95 eV, which fulfils the condition [42]. But the observed absorptive nonlinearity coefficient $β_{eff}$ is fairly high, which indicates that the observed RSA mechanism is mostly a result of ESA. The variations in the absorption cross-sections of excited state and ground state can give rise to a transition from RSA to SA mechanism [44,45]. The reduction in the PL emission intensity upon e-beam ex-posure underlies the quenching of defect centers and results in the unavailability of unoccupied defect states [44,45]. As a result the ground state absorption cross section is condensed, giving rise to the increase in transmittance towards the focus which results in the SA behavior.

Fig. 9(a) and (b) show open aperture Z-scan traces measured for $Mn_xZn_{1-x}O$ (x = 0.03, 0.05) nanostructures at different irradiation do-sages. A Q-switched Nd-YAG laser (wavelength 532 nm, pulse energy 100 μJ, pulse width 5ns) was employed for the measurement. It is ob-served that the transmittance curve exhibits a clear valley at the focus which can be attributed to RSA, indicating positive absorption non-linearity. A combination of ESA and TPA can be a more probable reason behind the presence of RSA mechanism in all the samples [46]. The most significant application of RSA is in optical limiting (OL) devices that shield sensitive optical components, including the human eye from laser-induced damage [47,48]. In addition, formation of defect states upon electron beam irradiation which results in the enhanced absorp-tion mechanism also results in prominent RSA mechanism exhibited by the nanostructures. The above mentioned mechanism can be explained as follows. The electrons in the ground state get excited to higher state by absorbing the photons from the incident laser intensity. The excited electrons can undergo different transition process. These excited elec-trons relax to the ground state in ns time scale by radiative and non-radiative decay. The excited electrons may also exhibit transitions to defect level by the mechanism of non-radiative relaxation within the ns time scale. The electrons in defect level can be further excited to the conduction band state which results in the process known as the excited state absorption (ESA) which in turn results in RSA mechanism ob-served in the investigated films [46].

From the Z-scan data obtained, the nonlinear absorption coefficient values $β_{eff}$ for both continuous and pulsed excitation regimes have been determined using the nonlinear transmission equations given by Sheik Bahae et al. [19,20], and the resulting data are shown in Table 5. Significant variations were observed in the nonlinear absorption coef-ficients for both continuous and pulsed laser regime upon electron beam irradiation. The enhancement observed in $β_{eff}$ on higher irra-diation dosages confirms the suitability of the investigated materials for optical limiting applications. The creation of intermediate states upon irradiation which further enhances electron excitation and relaxation processes in the materials could be the possible reason behind the ob-served variation in the magnitude of $β_{eff}$.

*3.5.2. Photo induced third harmonic generation*

Photo induced third harmonic generation (THG) measurement technique, may be considered as one of the most instructive technique for the precise assessment of third order nonlinearity raised in the materials due to electronic contribution. Generally the nonlinearity due to thermal, orientational, mechanical effects etc. can be neglected by implementing this technique [49]. The fundamental excitation beam used for the measurement possessed pulse duration was equal to 8 ns. (Nd: YAG pulsed laser with a wavelength of 1064 nm and frequency repetition rate of 10 Hz. The films were photo treated with a CW green laser light of wavelength 532 nm. Fig. 10(a) and (b) shows the photo induced THG signals of electron beam irradiated $Mn_xZn_{1-x}O$ (x = 0.03, 0.05) nanostructures. The value of the energy power for fundamental source beam have been varied from 140 J/m$^2$ to 200 J/m$^2$ by Glan polarizer and it was recorded by germanium photodetector supplied with defocused lens. THG signal observed in the films were detected by Hamamatsu photomultiplier by the aid of a filter of 355 nm. From the THG signal intensity vs laser power density graph shown in Fig. 10 it is observed a linear increase in the THG signal intensity on higher laser power density. To eliminate the parasitic background additional filters at 346 nm ad 460 nm have been used. The THG signal was at least one order higher with respect to the fluorescence parasitic scattering background. The measurements were performed after 0.1 s illumination by cw laser to avoid overheating in the 30 different points of the samples to achieve the necessary statistics. For $Mn_{0.03}Zn_{0.97}O$ and $Mn_{0.05}Zn_{0.95}O$ the films irradiated with 10 kGy e-beam dosage shows the highest intensity signal compared to other dosages. The substantial increment observed in the THG signal intensity indicates the credibility of electron beam irradiation in tailoring the THG signal intensity which can be used for the fabrication of frequency tripplers in high power laser sources. The observed increase in efficiency of photo induced THG in $Mn_xZn_{1-x}O$ (x = 0.03, 0.05) nanostructures can be possibly due to electron beam induced relaxation and photoexcitation process arised [50].

**4. Conclusions**

In summary, a series of $Mn_xZn_{1-x}O$ (x = 0.03, 0.05) nanostructures have been grown via the solution based chemical spray pyrolysis technique. Electron beam induced modifications on structural, optical and surface morphological properties have been studied and elabo-rated. Detailed investigations on the effect of electron beam irradiation on third order nonlinear optical properties are being reported for the first time. GXRD pattern shows sharp diffraction peaks matching with the hexagonal wurtzite structure of ZnO thin films. Gaussian decon-volution on PL spectra reveals the quenching of defect centers, implying the role of electron beam irradiation in regulating luminescence and

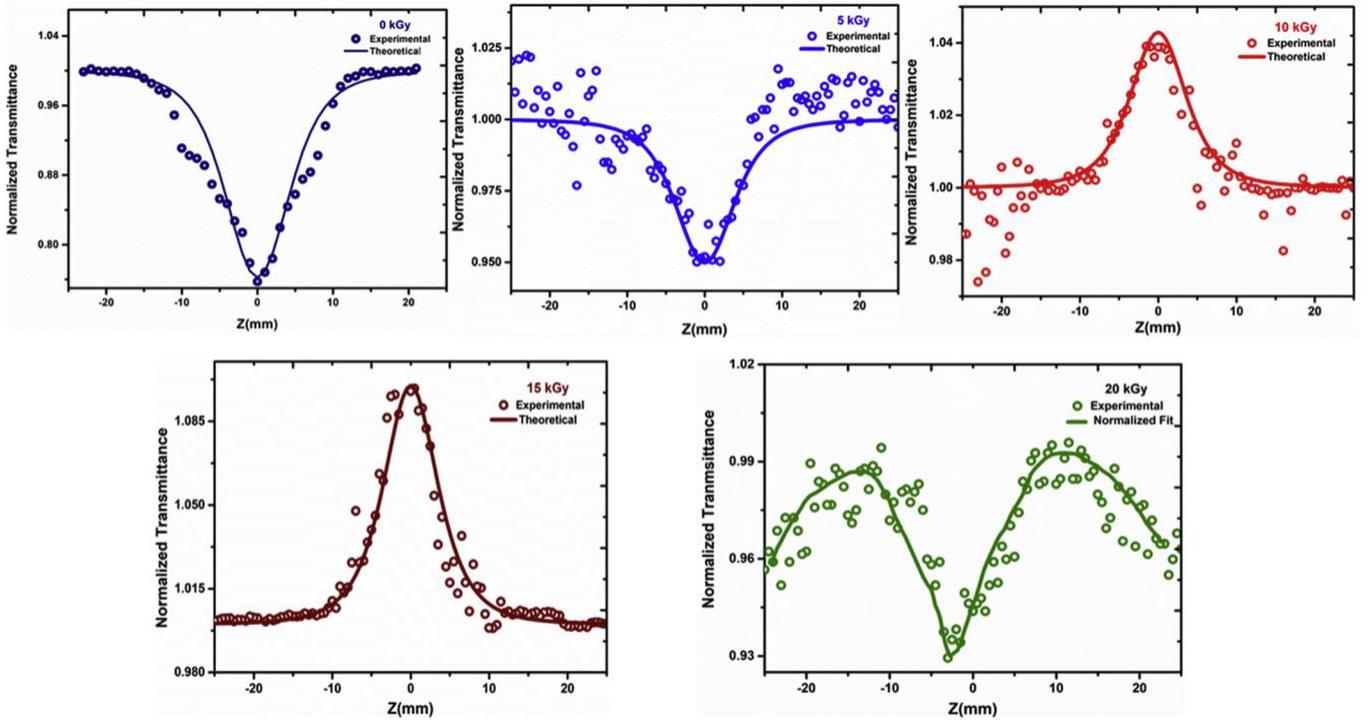

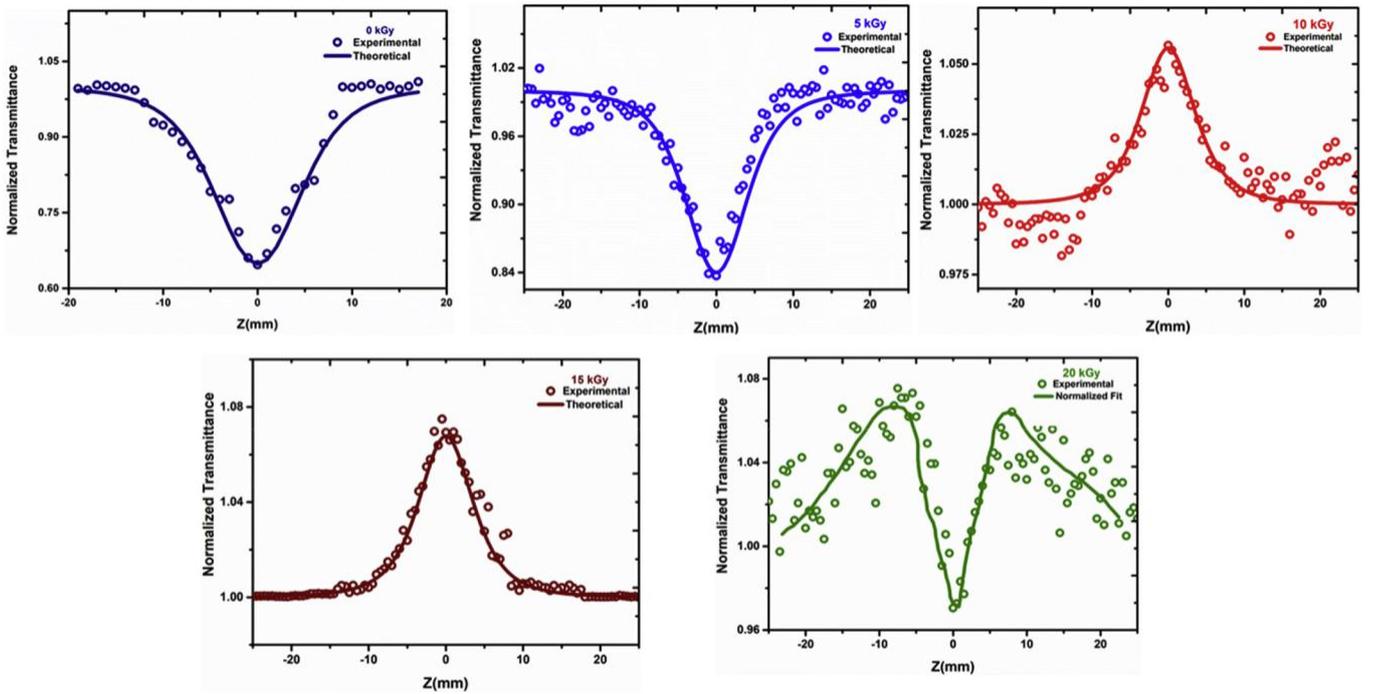

**Fig. 8.** Open aperture Z-scan traces measured for (a) $Mn_{0.03}Zn_{0.97}O$ (b) $Mn_{0.05}Zn_{0.95}O$ under continuous laser excitation.

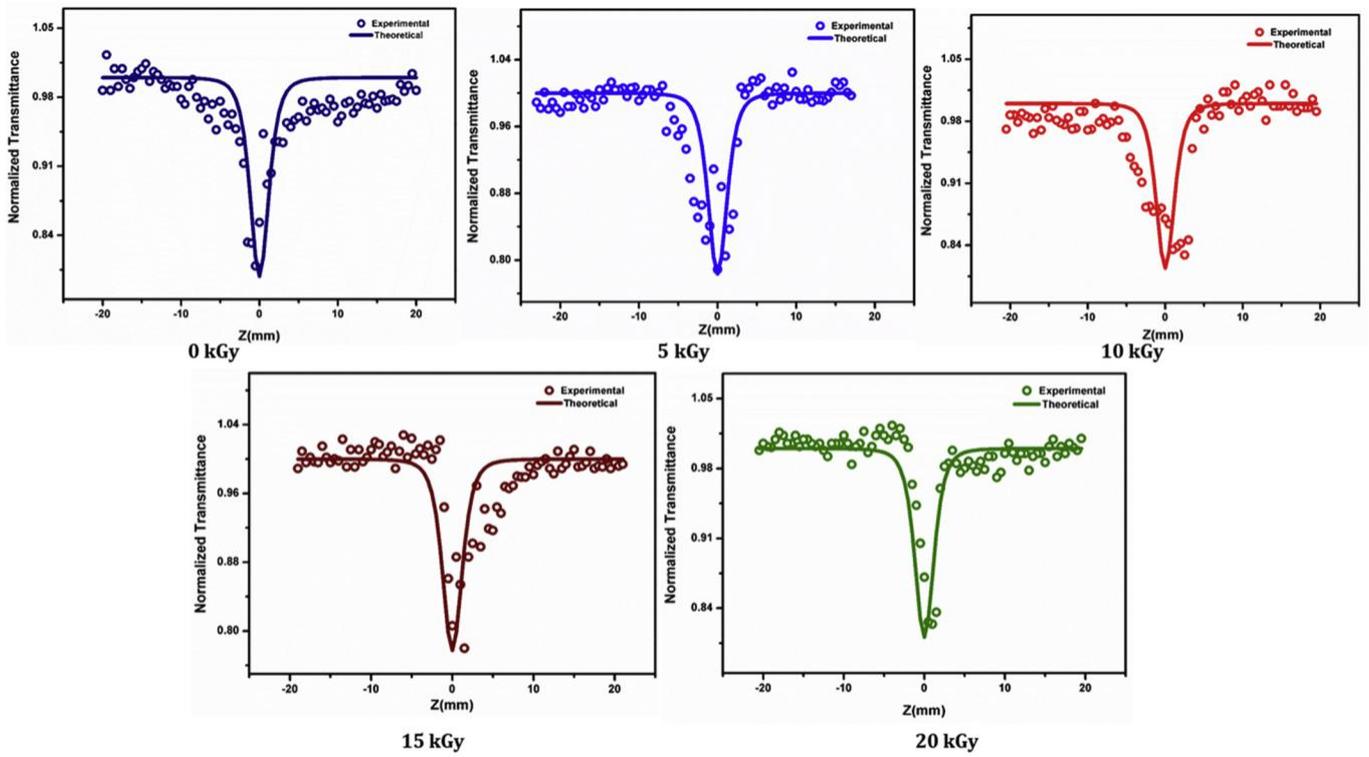

(a)

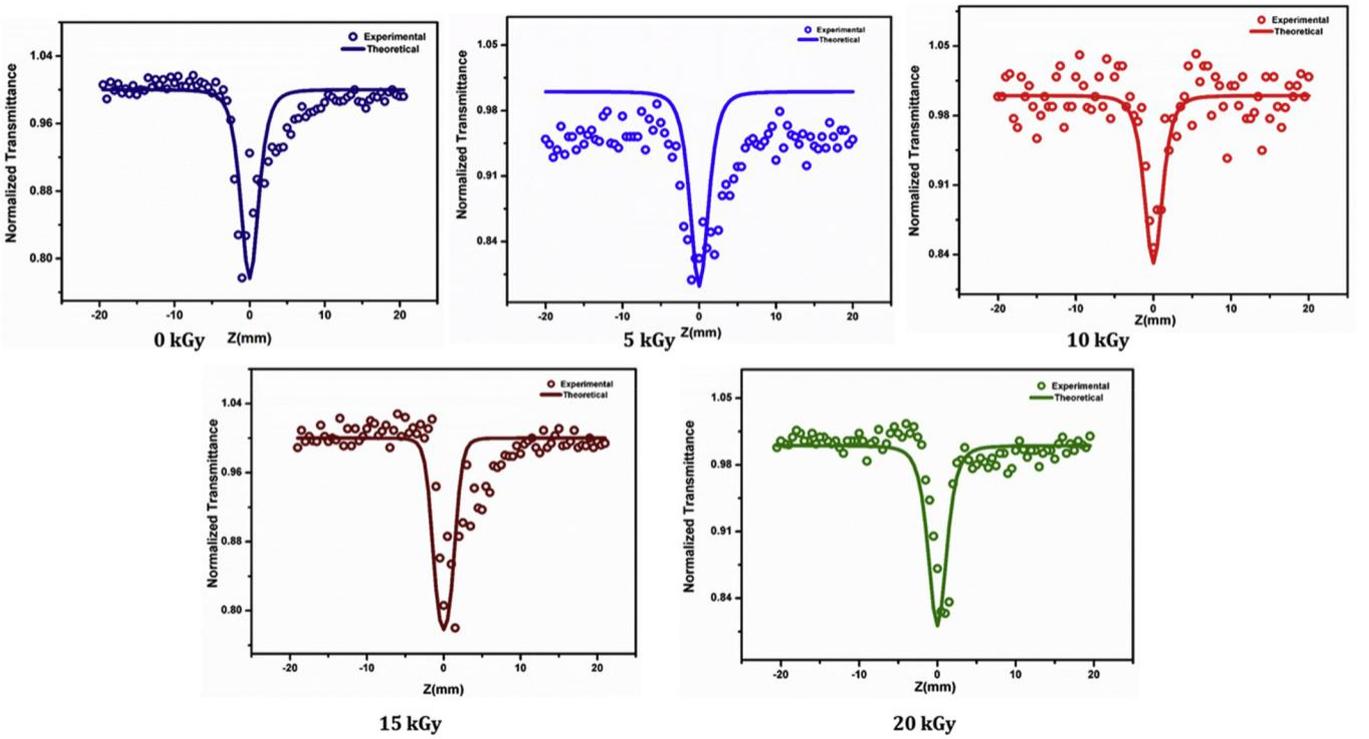

(b)

**Fig. 9.** Open aperture Z-scan curves measured of (a) $Mn_{0.03}Zn_{0.97}O$ (b) $Mn_{0.05}Zn_{0.95}O$ under pulsed laser excitation.

**Table 5**
Variation in nonlinear optical absorption coefficient in continuous and pulsed laser excitations upon electron beam irradiation.

| Sample | Laser Regime | Nonlinear Absorption Coefficient | Electron beam dosage(kGy) | | | | |
|---|---|---|---|---|---|---|---|
| | | | 0 | 5 | 10 | 15 | 20 |
| $Mn_{0.03}Zn_{0.97}O$ | Continuous | $\beta_{eff}$ (Cm/W) × $10^{-2}$ | 0.21 | 3.8 | −2.9 | −9.8 | 7.1 |
| | Pulsed | $\beta_{eff}$ (Cm/W) × $10^{-4}$ | 4.9 | 5.4 | 4.4 | 5.6 | 9.0 |
| $Mn_{0.05}Zn_{0.95}O$ | Continuous | $\beta_{eff}$ (Cm/W) × $10^{-2}$ | 0.31 | 14.3 | −3.7 | −7.9 | 2.9 |
| | Pulsed | $\beta_{eff}$ (Cm/W) × $10^{-4}$ | 9.4 | 8.5 | 9.7 | 9.3 | 9.0 |

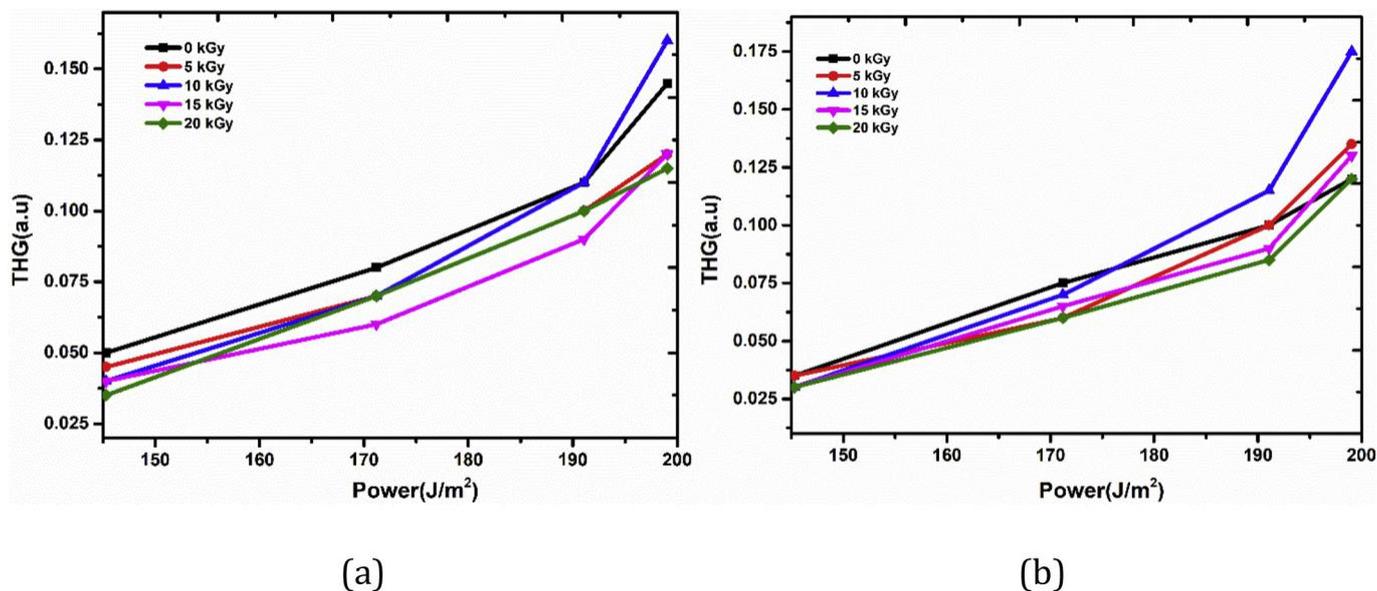

(a)      (b)

**Fig. 10.** THG versus fundamental power energy for (a) $Mn_{0.03}Zn_{0.97}O$ (b) $Mn_{0.05}Zn_{0.95}O$ irradiated at different e-beam dosages.

defect centers in the sample. Non-radiative transitions and structural disorder in $Mn_xZn_{1-x}O$ (x = 0.03, 0.05) thin films have been in-vestigated by Raman spectroscopy in the back scattering geometry. Irradiation induced spatial confinement and phonon localization effects have been observed in the sample, which are evident from spectral peak shifts and broadening. Third order absorptive nonlinearity of the sam-ples has been studied using the open aperture Z-scan technique. The measurements have been carried out in both continuous and pulsed laser excitation regimes. The strong RSA behavior shown by the sam-ples indicates their suitability for passive optical limiting applications. Laser induced third harmonic generation in the films shows substantial variations upon electron beam irradiation, and the effect can be utilized for frequency conversion of high power laser sources.

**Acknowledgements**

A part of this research was performed using facilities at CeNSE, funded by Ministry of Electronics and Information technology, Govt. of India, and located at the Indian Institute of Science, Bengaluru. For I. Kityk, J. Jedryka, K.Ozga this work was supported by a project European Union's Horizon 2020 research and innovation program under the Marie Skłodowska-Curie grant agreement No 778156.

**References**

[1] H. Shen, M. Wraback, J. Pamulapati, S. Liang, C. Gorla, Y. Lu, Properties of epitaxial ZnQ thin films for GaN and related application, MRS Internet J. Nitride Semicond. Res. 4S1 (1999) G3.60.
[2] M.J. Fejer, Nonlinear-optical frequency-conversion, Phys. Today 47 (1994) 25.
[3] A. Andrews, A. Schepartz, J. Sweedler, P. Weiss, Chemistry and the BRAIN in-itiative, J. Am. Chem. Soc. 136 (1) (2014) 1–2.
[4] P. Weiss, Where are the products of nanotechnology? ACS Nano 9 (2015) 3397.
[5] B. Pelaz, S. Jaber, D. de Aberasturi, V. Wulf, T. Aida, J. de la Fuente, J. Feldmann, H. Gaub, L. Josephson, C. Kagan, N. Kotov, L. Liz-Marz´an, H. Mattoussi, P. Mulvaney, C. Murray, A. Rogach, P. Weiss, I. Willner, W. Park, The state of nanoparticle-based nanoscience and nanobiotechnology: progress, promises, and chal-lenges, ACS Nano 6 (2012) 8468.
[6] Anton Autere, Henri Jussila, Yunyun Dai, Yadong Wang, HarriLipsanen, Zhipei Sun, Nonlinear optics with 2D layered materials, Adv. Mater. (2018) 1705963.
[7] Y.R. Shen, The Principles of Nonlinear Optics, Wiley Press, NewYork, 1984.
[8] E. Garmire, Nonlinear optics in daily life, Optic Express 21 (25) (2013) 30532–30544.
[9] G.P. Agrawal, Applications of Nonlinear Fiber Optics, Academic Press, London, UK, 2001.
[10] B.E.A. Saleh, M.C. Teich, Fundamentals of Photonics, Wiley, Hoboken, NJ, USA, 2007.
[11] R.W. Boyd, Nonlinear Optics, Academic Press, New York, 2007.
[12] G.I. Petrov, V. Shcheslavskiy, V.V. Yakovlev, I. Ozerov, E. Chelnokov, W. Marine, Efficient third-harmonic generation in a thin nanocrystalline film of ZnO", Appl. Phys. Lett. 83 (2003) 3993.
[13] Albin Antony, P. Poornesh, K. Ozga, J. Jedryka, P. Rakus, I.V. Kityk, Enhancement of the efficiency of the third harmonic generation process in ZnO:F thin films probed by photoluminescence and Raman spectroscopy", Mater. Sci. Semicond. Process. 87 (2018) 100–109.
[14] Pairot Moontragoon, Supree Pinitsoontorn, Prasit Thongbai, Mn-doped ZnO nano-particles: preparation, characterization, and calculation of electronic and magnetic properties, Microelectron. Eng. 108 (2013) 158–162.
[15] A.H. Reshak, K. Nouneh, I.V. Kityk, J. Bila, S. Auluck, H. Kamarudin, Z. Sekkat, Structural, electronic and optical properties in earth abundant photovoltaic ab-sorber of Cu2ZnSnS4 and Cu2ZnSnSe4 from DFT calculations, Int. J. Electrochem. Sci. 9 (2014) 955.
[16] I.V. Kityk, J. Ebothe, AelHichou, M. Addou, A. Bougrine, B. Sahraoui, "Linear electro-optics effect in ZnO-F film-glass interface", Phys. Status Solidi 234B (2002) 553.
[17] H.A. Khawal, B.N. Dole, "A study of the 160 MeV Ni7+swift heavy ion irradiation effect of defect creation and shifting of the phonon modes on $MnxZn1–xO$ thin films", RSC Adv. 7 (2017) 34736.
[18] S.K. Neogi, S. Chattopadhyay, Aritra Banerjee, S. Bandyopadhyay, A. Sarkar, Ravi Kumar, Effect of 50 MeV Li3 + irradiation on structural and electrical prop-erties of Mn-doped ZnO, J. Phys. Condens. Matter 23 (20) (2011) 5801.
[19] M. Sheik-Bahae, A.A. Said, E.W. Van Stryland, High-sensitivity, single-beam n2 measurements, Opt. Lett. 14 (1989) 955.
[20] M. Sheik-Bahae, A.A. Said, T.H. Wei, D.J. Hagan, E.W. Van Stryland, Sensitive

measurement of optical nonlinearities using a single beam IEEE, J. Quantum Elect. 26 (1990) 760.

[21] M. Ghotbi, Z. Sun, A. Majchrowski, E. Michalski, I.V. Kityk, M. Ebrahim-Zadeh, Efficient third harmonic generation of microjoule picosecond pulses at 355nm in BiB3O6", Appl. Phys. Lett. 89 (2006) 173124.

[22] V.V. Ison, A. Ranga Rao, V. Dutta, P.K. Kulriya, D.K. Avasthi, S.K. Tripathi, Swift heavy ion induced structural changes in CdS thin films possessing different microstructures: a comparative study, J. Appl. Phys. 106 (2009) 023508.

[23] J. Tauc, R. Grigorovici, A. Vancu, Optical properties and electronic structure of amorphous germanium, Phys. Status Solidi 15 (1966) 627.

[24] S. Som, Subrata Das, S. Dutta, Mukesh Kumar Pandey, Ritesh Kumar Dubey, H.G. Visser, S.K. Sharma, S.P. Lochab, A comparative study on the influence of 150 MeV Ni7+, 120 MeV Ag9+, and 110 MeV Au8+ swift heavy ions on the structural and thermoluminescence properties of Y2O3: Eu3+/Tb3+ nanophosphor for do-simetric applications, J. Mater. Sci. 51 (2016) 1278.

[25] Jitendra Pal Singh, I. Sulania, Jai Prakash, S. Gautam, K.H. Chae, D. Kanjilal, K. Asokan, Study of surface morphology and grain size of irradiated MgO thin films, Adv. Mat. Lett. 3 (2) (2012) 112.

[26] A.B. Djurisic, A.M.C. Ng, X.Y. Chen, ZnO nanostructures for optoelectronics: ma-terial properties and device applications, Prog. Quant. Electron. 34 (2010) 191.

[27] Debajyoti Das, PraloyMondal, Photoluminescence phenomena prevailing in c-axis oriented intrinsic ZnO thin films prepared by RF magnetron sputtering", RSC Adv. 4 (2014) 3573.

[28] S.A.M. Lima, F.A. Sigoli, M. JafelicciJr, M.R. Davolos, Luminescent properties and lattice defects correlation on zinc oxide, Int. J. Inorg. Mater. 3 (2001) 749.

[29] N.H. Nickel, E. Terukov, Zinc Oxide A Material for Micro-and Optoelectronic Applications, Springer, Dodrecht, The Netherlands, 2005, p. 69.

[30] J. Hu, B.C. Pan, Electronic structures of defects in ZnO: hybrid density functional studies, J. Chem. Phys. 129 (15) (2008) 154706.

[31] N. Midya, a S.K. Neogi, b Md A. Ahmed, a A. Banerjee, d D. Kanjilald ac Pravin Kumar, S. Bandyopadhyay, Correlation between magnetic and micro-structural properties of low energy ion irradiated and un-irradiated Zn0.95Mn0.05O films", RSC Adv. 7 (2017) 771.

[32] L. Duan, G. Rao, Y. Wang, J. Yu, T. Wang, Magnetization and Raman scattering studies of (Co,Mn) codoped ZnO nanoparticles, J. Appl. Phys. 104 (2008) 013909.

[33] M. Koyano, PhungQuocBao, Le thiThanhBinh, HongHa Le, NgocLong Nguyen, Katayama Shin'ichi, Photoluminescence and Raman spectra of ZnO thin films by charged liquid cluster beam technique, phys. stat. sol. 193 (No. 1) (2002) 125–131.

[34] Khalid, M.A. Malik, D. Lewis, P. Kevin, E. Ahmed, Y. Khan, P. O'Brien, Transition metal doped pyrite (FeS2) thin films: structural properties and evaluation of optical band gap energies, J. Mater. Chem. C 3 (2015) 12068.

[35] Shuxia Guo, Zuliang Du, Shuxi Dai, Analysis of Raman modes in Mn-doped ZnO nanocrystals, Phys. Status Solidi b 246 (10) (2009) 2329–2332.

[36] Albin Antony, P. Poornesh, I.V. Kityk, K. Ozga, Ganesh Sanjeev, Vikash Chandra Petwal, Vijay Pal Verma, Jishnu Dwivedi, A novel approach for tailoring structural, morphological, photoluminescence and nonlinear optical features in spray coated Cu:ZnO nanostructures via e-beam", CrystEngComm 20 (2018) 6502.

[37] R. Baghdad, B. Kharroubi, A. Abdiche, M. Bousmaha, M.A. Bezzerrouk, A. Zeinert, M. E Marssi, K. Zellama, Mn doped ZnO nanostructured thin films prepared by ultrasonic spray pyrolysis method, Superlattice. Microst. 52 (2012) 711.

[38] Li Jin, Huiqing Fan, Xiaopeng Chen, Zhiyi Cao, Structural and photoluminescence of Mn-doped ZnO single-crystalline nanorods grown via solvothermal method, Colloid. Surface. Physicochem. Eng. Aspect. 349 (2009) 202.

[39] A. Ali Fatima, Suganthi Devadason, T. Mahalingam, Structural, luminescence and magnetic properties of Mn doped ZnO thin films using spin coating technique", J. Mater. Sci. Mater. Electron. 25 (2014) 3466.

[40] H. Richter, Z.P. Wang, L. Ley, The one phonon Raman spectrum in microcrystalline silicon, Solid State Commun. 39 (1981) 625.

[41] I.H. Campbell, P.M. Fauchet, The effects of microcrystal size and shape on the one phonon Raman spectra of crystalline semiconductors, Solid State Commun. 58 (10) (1986) 739.

[42] K.K. Nagaraja, S. Pramodini, A. Santhosh Kumar, H.S. Nagaraja, P. Poornesh, DhananjayaKekuda, "Third-order nonlinear optical properties of Mn doped ZnO thin films under cw laser illumination", Opt. Mater. 35 (2013) 431–439.

[43] M. Abd-Lefdil, A. Douayar, A. Belayachi, A.H. Reshak, A.O. Fedorchuk, S.Pramodini, P. Poornesh, K.K. Nagaraja, H.S. Nagaraja, Third harmonic generation process in Al doped ZnO thin films, J. Alloy. Comp. 584 (2014) 7.

[44] J. Wang, F. Jin, X. Cao, S. Cheng, C. Liu, Y. Yuan, J. Fang, H. Zhao, J. Li, In2Te3 thin films: a promising nonlinear optical material with tunable nonlinear absorption response, RSC Adv. 6 (2016) 103357.

[45] A. Singh, S. Kumar, R. Das, P.K. Sahoo, Defect-assisted saturable absorption char-acteristics in Mn doped ZnO nano-rods, RSC Adv. 5 (2015) 88767.

[46] Ke-Xin Zhang, Cheng-Bao Yao, Xing Wen, Qiang-Hua Li, Wen-Jun Sun, Ultrafast nonlinear optical properties and carrier dynamics of silver nanoparticle-decorated ZnO nanowires, RSC Adv. 8 (2018) 26133.

[47] Rajeswari Ponnusamy, Dhanuskodi Sivasubramanian, P. Sreekanth, Vinitha Gandhiraj, Reji Philip, G.M. Bhalerao, Nonlinear optical interactions of Co: ZnO nanoparticles in continuous and pulsed mode of operations, RSC Adv. 5 (2015) 80756–80765.

[48] Benoy Anand, S.R. Krishnan, Ramakrishna Podila, S. Siva Sankara Sai, Apparao M. Rao, Reji Philip, The role of defects in the nonlinear optical absorption behavior of carbon and ZnO nanostructures, Phys. Chem. Chem. Phys. 16 (2014) 8168.

[49] L. Castaneda, O.G. Morales-Saavedra, D.R. Acosta, A. Maldonado, "Structural, morphological, optical, and nonlinear optical properties of fluorine☐doped zinc oxide thin films deposited on glass substrates by the chemical spray technique", Phys. Stat. Sol. 203 (2006) 1971.

[50] M. Abd-Lefdil, A. Belayachi, S. Pramodini, P. Poornesh, A. Wojciechowski, A.O. Fedorchuk, "Structural, photoinduced optical effects and third-order nonlinear optical studies on Mn doped and Mn–Al codoped ZnO thin films under continuous wave laser irradiation", Laser Phys. 24 (2014) 035404.